\documentclass[a4paper,12pt]{article}
\usepackage[english]{babel}
\usepackage[a4paper,tmargin=3truecm,bmargin=3truecm,rmargin=2.5truecm,lmargin=2.5truecm,twoside,verbose=true]{geometry}

\usepackage{fabmacromath}

\usepackage{graphicx}
\usepackage{cancel}

\usepackage{slashed}
\usepackage{amsmath,amssymb}

\numberwithin{equation}{section}

\allowdisplaybreaks[1]

\title{One-loop Beta Functions for the Orientable\\Non-commutative Gross-Neveu Model\footnote{Work supported by ANR grant NT05-3-43374 ``GenoPhy''.}}
\author{Ahmed Lakhoua$^a$, Fabien Vignes-Tourneret$^b$, Jean-Christophe Wallet$^a$}
\date{}

\begin{document}

\maketitle
\vspace*{-1cm}
\begin{center}
\textit{$^a$Laboratoire de Physique Th\'eorique, B\^at.\ 210\\
    Universit\'e Paris XI,  F-91405 Orsay Cedex, France\\
    e-mail: \texttt{jean-christophe.wallet@th.u-psud.fr}}\par
\textit{$^b$IH\'ES, Le Bois-Marie, 35 route de Chartres, F-91440 Bures-sur-Yvette, France\\
    e-mail: \texttt{vignes@ihes.fr}}\
 \end{center}%

\begin{abstract}
We compute at the one-loop order the $\beta$-functions for a renormalisable non-commutative analog of the Gross Neveu model defined on the Moyal plane. The calculation is performed within the so called $x$-space formalism. We find that this non-commutative field theory exhibits asymptotic freedom for any number of colors. The $\beta$-function for the non-commutative counterpart of the Thirring model is found to be non vanishing. 
\end{abstract}%
\pagebreak
\section{Introduction}
In the past few years, some activity has been focused on the study of various classes of field theories defined on Moyal spaces \cite{REVIEW}. These provide prototypes of non-commutative field theories which are interesting in themselves since they involve some salient features of non-commutative geometry \cite{CONNES}. The interest in the study of these field theories was further increased by the claim that somehow similar non-commutative field theories seem to emerge rather naturally from (some limiting regime of) String Theory and matrix theory in magnetic backgrounds \cite{STRINGERY}. Recall that in non-commutative geometry, the commutative algebras of functions defined on differentiable manifolds (roughly the coordinates spaces) are replaced by associative but non-commutative algebras further interpreted as algebras of functions on ``non-commutative spaces''. Within this later algebraic framework, natural non-commutative analogs of the main geometrical objects usually involved in field theories can be algebraically defined, such as for instance connections, curvatures, vector bundles, so that the construction of various non-commutative analogs of fields theories can be undertaken. The starting relevant configuration spaces for the non-commutative field theories are modules over the associative algebras which are naturally viewed as non commutative analogs of vector bundles. One example of associative algebra among many others is provided by the associative Moyal algebras \cite{Gracia-Bondia1987kw,Varilly1988jk} therefore playing the role of ``non-commutative Moyal spaces''. At this level, one technical remark is in order. Throughout this paper, we will consider only ${\mathbb{R}}_\theta^D$ algebras, i.e the non-commutative counterpart of the {\it{Euclidean}} ${\mathbb{R}}^D$ spaces. For studies related to non-commutative tori, see e.g. \cite{REVIEW} and references therein. Besides, the ensuing discussion refers to non-commutative field theories defined on free modules. \par
The simplest non-commutative generalizations on Moyal spaces of the usual scalar theories that have been first investigated were shown to suffer from the so called UV/IR mixing \cite{VANRAMS}, a phenomenon that makes the renormalisability very unlikely. Recall that UV/IR mixing results from the existence of potentially dangerous non-planar diagrams which, albeit UV finite, become singular at exceptional (low) external momenta. This triggers the occurrence of UV divergences in higher order diagrams in which they are involved as subdiagrams. This signals that UV and IR scales are related in a non trivial way which should in principle invalidate a Wilson-type renormalisation scheme \cite{WILSONSCH}. An appealing solution to the UV/IR mixing has been recently proposed  by Grosse and Wulkenhaar \cite{GrWu03-1} within the  non-commutative $\varphi^4$ model on the 4-dimensional Moyal space where $\varphi$ is real-valued. They showed that the UV/IR mixing can be suppressed by supplementing the initial action with a harmonic oscillator quadratic term leading to a renormalisable non-commutative quantum field theory. The initial proof \cite{GrWu03-1} was performed within the matrix-base formalism, roughly a basis for the (Schwarz class) functions for which the associative product of the Moyal algebra is a simple matrix product. This cumbersome proof was simplified through a reformulation into the (position) $x$-space formalism in \cite{SIMPLIF} which exhibits some advantages compared to the matrix-base formulation. For instance, the propagator in $x$-space can be explicitely computed (as a Mehler kernel \cite{RIVASSEAUPROPA} ) and {\it{actually}} used in calculations. Besides, it makes easier the comparition of the renormalisation group for non-commutative theories and their commutative counterpart.\par
At the present time, another renormalisable non-commutative scalar quantum field theory on Moyal spaces has been identified. This is a (complex-valued) scalar theory studied in \cite{SIMPLIF} which can be viewed as a modified version of the LSZ model \cite{LSZ-1} (the scalar theory in \cite{SUPERRENORM} is super renormalisable). Note that interesting solvable non-commutative scalar field theories have also been considered in \cite{LSZBIS}. As far as Fermionic theories are concerned, a non-commutative version on the Moyal plane of the Gross-Neveu model \cite{GROSSNEVEU}, called the orientable non-commutative Gross Neveu model,  has been recently considered and shown to be renormalisable to all orders \cite{RenNCGN05,vignes-tourneret06:PhD}. It is worth mentioning that this non-commutative field theory still exhibits some UV/IR mixing, even in the presence of the Fermionic version of the harmonic oscillator quadratic term introduced in \cite{GrWu03-1}, which however does not prevent the theory to be renormalisable. Note that in \cite{SEMENOFF} the large $N$ limit of the non-commutative Gross-Neveu model, with however a restricted interaction, has been studied; renormalisability is shown at this limit together with asymptotic freedom. One should keep in mind that the fact that the orientable Gross Neveu model is renormalisable although it involves some remaining UV/IR mixing \cite{RenNCGN05} indicates that further investigations are needed to actually clarify the effective role of the various generalizations of the above mentionned harmonic oscillator term and of the related covariance of the considered theory under the Langmann-Szabo duality \cite{LANGSZAB} and their impact in the control of the UV/IR mixing and renormalisability.\par
Despite this remaining uncertainty, coupling constant flows and $\beta$-functions can be studied in the available renormalisable non-commutative field theories. The $\beta$-function for the coupling constant of the non-commutative (real-valued) $\varphi^4$ model on the 4-dimensional Moyal space has been computed at the one-loop order in \cite{GrWu04-2}. It exhibits a bounded flow, finite fixed point and vanishes when the parameter affecting the harmonic oscillator quadratic term, says $\Omega$,  is equal to unity, which corresponds to the (self-dual) point in the parameter space where the field theory is invariant under the Langmann-Szabo duality \cite{LANGSZAB}. This latter result has been proven very recently to be valid to all orders in \cite{DISSERT}. In the present paper, we compute at the one-loop order the $\beta$-functions of the coupling constants involved in the renormalisable non-commutative version on ${\mathbb{R}}^2_\theta$ of the Gross-Neveu model considered in \cite{RenNCGN05}. Recall that the (commutative) Gross-Neveu model exhibits asymptotic freedom together with mass generation phenomenon \cite{GROSSNEVEU}.
In section 2, we recall the principal features of the renormalisable non-commutative version of the Gross-Neveu model and collect the various ingredients relevant for the calculation. The section 3 is devoted to the one-loop computation of the relevant correlation functions. The analysis is carried out within the $x$-space formalism which appears to be well adapted for the relevant calculations.  In section 4, we present the expressions for the $\beta$-functions, collect and discuss the main results and finally we conclude.\par

\section{The Orientable non-commutative Gross-Neveu Model}
In this section we recall the main features of the relevant action \cite{RenNCGN05,vignes-tourneret06:PhD}, fix the conventions and collect the useful material that will be needed throughout this paper. Let $\mathbb{R}^2_\theta$ denotes the ``Moyal plane'' \cite{Gracia-Bondia1987kw,Varilly1988jk}, which can be viewed in the following as a unital involutive associative algebra over $\mathbb{C}$ generated by the coordinate functions on $\mathbb{R}^2$ such that $[x^\mu,x^\nu]_\star=\imath\Theta^{\mu\nu}$ with $[a,b]_\star= a\star b-b\star a$ for any $(a,b)\in\mathbb{R}^2_\theta$. Here, ``$\star$'' denotes the associative Moyal-Groenwald product on $\mathbb{R}^2_\theta$ induced by $\Theta$, an invertible constant skew-symmetric matrix, that can be chosen to be $\Theta=\theta{\mathcal{S}}$, ${\mathcal{S}}=\lbt\begin{smallmatrix} 0 & -1 \\ 1 & \phantom{-}0 \end{smallmatrix}\rbt$ where the parameter $\theta$ has mass dimension $-2$. The Moyal product can be represented as
\begin{align}
  (a\star b)(x)=&\frac{1}{(2\pi)^2}\int d^2yd^2k\, a(x+{\textstyle\frac 12}\Theta.k)b(x+y)e^{ik.y}\label{eq:MoyalProd}
\end{align}
where $(\Theta.k)^{\mu}=\Theta^{\mu\nu}k_{\nu}$. We also define $X.\Theta^{-1}.Y=X^\mu\Theta^{-1}_{\mu\nu}Y^\nu$. For more mathematical details, see e.g \cite{Gracia-Bondia1987kw,Varilly1988jk}.\par
The action for the orientable non-commutative Gross-Neveu model on $\mathbb{R}^2_\theta$ \cite{RenNCGN05} can be written as
\begin{subequations}
  \begin{align}
    S=&\int d^2x\big[{\bar{\psi}}(-i{\slashed{\partial}}+\Omega\xts+m+\kappa\gamma_5)\psi-
    \sum_{A=1}^3\frac{g_A}{4}({\cal{J}}^{A}\star{\cal{J}}^{A})(x)\big]\label{eq:action}\\
    {\cal{J}}^{A}=&{\bar{\psi}}\star\Gamma^{A}\psi,\ \Gamma_{1}=\bbbone,\,\Gamma_{2}=\gamma^{\mu},\,\Gamma_{3}=\gamma_{5}\label{eq:courants}
  \end{align}
\end{subequations}
where $\slashed{a}=a_\mu\gamma^\mu$, $\xt=2\Theta^{-1}.x$ and a summation over the Lorentz indices $\mu$ is understood in the interaction term involving ${\cal{J}}_2$,  ${\cal{J}}_2\star{\cal{J}}_2$$=$$\sum_\mu({\bar{\psi}}\star\gamma_\mu\psi)\star({\bar{\psi}}\star\gamma^\mu\psi)$. The Clifford algebra for the 2D anti-Hermitian gamma matrices satisfy $\{\gamma^{\mu},\gamma^{\nu}\}=-2\delta^{\mu\nu}$ and $\gamma_5=\imath\gamma^{0}\gamma^{1}$. The field $\psi$ denotes a $2N$-component spinor field where $N$ is the number of \emph{colors}. The parameters $\Omega$ (to be discussed in a while), $0\le\Omega<1$,  and the $g_A$'s are dimensionless. In \eqref{eq:action}, the term $\Omega{\bar{\psi}}\xt\psi$ can be viewed as the Fermionic counterpart of the harmonic oscillator term first introduced in \cite{GrWu03-1}. Here, two comments are in order. First, notice that the minus sign affecting the four-Fermion interaction term in \eqref{eq:action} is a mere generalization of the interaction term in the commutative Gross Neveu model \cite{GROSSNEVEU} for which asymptotic freedom is obtained. Next, the fact that the model defined in \eqref{eq:action} is called ''orientable'' comes from the present choice for the interactions. Recall that within the present non-commutative framework, six independent four-fermion interactions can in principle be constructed. The three interaction terms in  which $\psi$ and ${\bar{\psi}}$ alternate, namely $\sum_{a,b}{\bar{\psi}}_a\star\psi_b\star{\bar{\psi}}_a\star\psi_b$, $\sum_{a,b}{\bar{\psi}}_a\star\psi_b\star{\bar{\psi}}_b\star\psi_a$, $\sum_{a,b}{\bar{\psi}}_a\star\psi_a\star{\bar{\psi}}_b\star\psi_b$ (the sum runs over color indices $a$, $b$), gives rise after suitable Fierz transformations to the interaction term in \eqref{eq:action} from which the diagrams occurring in the loopwise expansion can be given an orientation \cite{RenNCGN05}. This explains why \eqref{eq:action} is called ''orientable''. Three other terms with adjacent $\psi$ and ${\bar{\psi}}$ could in principle be written, namely $\sum_{a,b}{\bar{\psi}}_a\star{\bar{\psi}}_b\star\psi_a\star\psi_b$, $\sum_{a,b}{\bar{\psi}}_a\star{\bar{\psi}}_b\star\psi_b\star\psi_a$, $\sum_{a,b}{\bar{\psi}}_a\star{\bar{\psi}}_a\star\psi_b\star\psi_b$. Such interaction terms would result in a field theory with diagrams that cannot be orientated (in addition to the orientable one) \cite{RenNCGN05}. Although the proper interpretation of the non-orientable interactions is not clear within the present algebraic framework, the corresponding field theory is interesting in itself. Its detailed study has been undertaken in \cite{NOUS}.\par
\begin{figure}[!htb]
  \centering
  \includegraphics[scale=1]{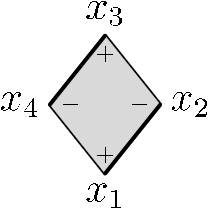}
  \caption[The Vertex]{\footnotesize{Graphical representation for the vertex in the $x$-space, obtained from \eqref{eq:interaction}. The 
  plus-sign 
  (resp.minus-sign) appearing in the rhombus corresponds to incoming (resp. outgoing) external line associated with ${\bar{\psi}}$ 
  (resp. $\psi$).}}
  \label{fig:vertex}
\end{figure}

The action \eqref{eq:action} has been shown to be renormalisable to all orders in \cite{RenNCGN05}. Notice that the proof relies rather heavily on the orientability of the diagrams. In the massive case ($m$$\neq$$0$), the term $\kappa{\bar{\psi}}\gamma_5\psi$ in \eqref{eq:action}, even not present at the classical level would be generated by higher order corrections (at the two-loop order) \cite{RenNCGN05}. The Feynman graphs can be computed from the propagator and interaction vertex derived from \eqref{eq:action}. In the following, we will work within the $x$-space formalism \cite{SIMPLIF} which proves convenient in the present analysis. The propagator in the $x$-space $C(x,y)=(-i{\slashed{\partial}}+\Omega\xts+m)^{-1}(x,y)$ can be written as \cite{toolbox05,vignes-tourneret06:PhD}
\begin{subequations}
  \label{eq:Propa}
  \begin{align}
    C(x,y)=&\int_0^\infty dtC(t;x,y);\ C(t;x,y)=\Cb(t;y-x)\exp(\imath\Omega x\wedge y),\label{eq:Cbar}
    \intertext{where $x\wedge y=2x.\Theta^{-1}.y$ and}
    \Cb(t;u)=&\frac{\Omega}{\pi\theta}\frac{e^{-m^2t}}{\sinh(2\Ot t)}e^{-\frac{\Ot}{2}\coth(2\Ot t)u^2}\big(\imath\Ot\coth(2\Ot t)\slashed{u}+
    \Omega\uts+m\big)e^{-2\gamma_5\Ot t},\label{eq:PropaX}
  \end{align}
\end{subequations}
with $\Ot=\frac{2\Omega}{\theta}$. Note that the following formula
\begin{align}
  \label{eq:ExpGamma5}
e^{-\alpha\gamma_5}=&\cosh(\alpha)\bbbone-\sinh(\alpha)\gamma_5
\end{align}
holds for any real parameter $\alpha$. The propagator $C$ is diagonal in its color indices. The interaction vertices can be read off from the RHS of
\begin{subequations}
  \label{eq:interaction}
  \begin{align}
    \int d^2x\,({\bar{\psi}}\star\Gamma_A\psi\star{\bar{\psi}}\star\Gamma_A\psi)=&\frac{1}{\pi^2\theta^2}\int 
    \prod_{i=1}^4d^2x_i\,{\bar{\psi}}(x_1)\Gamma_A\psi(x_2){\bar{\psi}}(x_3)\Gamma_A\psi(x_4)\label{eq:InteractionDvpee
    }\\
    &\times\delta(x_1-x_2+x_3-x_4)e^{-\imath\sum_{i<j}(-1)^{i+j+1}x_i\wedge x_j }.\nonumber
    \intertext{We will denote the vertex kernel as}
    V(x_1,x_2,x_3,x_4)=&\delta(x_1-x_2+x_3-x_4)e^{-\imath\sum_{i<j}(-1)^{i+j+1}x_i\wedge x_j }.\label{eq:InteractionKernel}
  \end{align}
\end{subequations}
The graphical representation of the vertex is depicted on the figure~\ref{fig:vertex}. The non-locality of the interaction is conveniently represented by the rhombus appearing on fig.~\ref{fig:vertex} whose vertices correspond to the $x_i$'s occurring in \eqref{eq:interaction}. It is usefull to represent the alternate signs in the delta function of \eqref{eq:interaction} by plus- and minus-signs, as depicted on the figure. By convention, a plus-sign (resp. minus-sign) corresponds to an incoming field ${\bar{\psi}}$ (resp. outgoing field $\psi$). This permits one to define an orientation on the diagrams obtained from the loop expansion. Notice that external lines are not drawn explicitly as we will deal essentially with amputated Green functions.

In the computation of the relevant diagrams, recall that a factor
\begin{align}
    \sum_{A=1}^3\frac{g_A}{4\pi^2\theta^2}\int {\cal{D}}(x);\ {\cal{D}}(x)=\prod_{i=1}^4 d^2x_i\, V(x_1,x_2,x_3,x_4)
\end{align}
must appear in the amplitude for each involved (square) vertex. Besides, the contraction between $\psi$ and ${\bar{\psi}}$ used in the computation of the amplitudes is defined by $C(x,y)=\int{\cal{D}}{\bar{\psi}}{\cal{D}}\psi e^{-S_{\text{free}}}\psi(x)\psib(y)$. Furthermore, the following formulas among the $\gamma$ matrices will be useful
\begin{subequations}
  \label{eq:GammaFormulae}
  \begin{align}
    \gamma^\mu\gamma^\nu=&-\delta^{\mu\nu}\bbbone-\imath\epsilon^{\mu\nu}\gamma_5&\gamma^\mu\gamma^\nu\gamma^\rho=&(\delta^{\mu\nu}\delta^{\rho}_{\phantom{\rho}\sigma}-\delta^{\mu\rho}\delta^{\nu}_{\phantom{\nu}\sigma}+\delta^{\nu\rho}\delta^{\mu}_{\phantom{\mu}\sigma})\gamma^\sigma\label{eq:Gamma}\\
    \gamma_5\gamma^\mu=&-\imath\epsilon^{\mu}_{\phantom{\mu}\nu}\gamma^\nu&\gamma_5\gamma^\mu\gamma^\nu=&-\imath\epsilon^{\mu\nu}\bbbone-\delta^{\mu\nu}\gamma_5\label{eq:Gamma5}\\
    \Tr(\gamma^\mu\gamma^\nu)=&-2\delta^{\mu\nu}&\Tr(\gamma^\mu\gamma^\nu\gamma^\rho\gamma^\sigma)=&2(\delta^{\mu\nu}\delta^{\rho\sigma}-\delta^{\mu\rho}\delta^{\nu\sigma}+\delta^{\mu\sigma}\delta^{\nu\rho})\label{eq:TrGamma}\\
    \Tr(\gamma_5\gamma^\mu)=&0&\Tr(\gamma_5\gamma^\mu\gamma^\nu)=&-2\imath\epsilon^{\mu\nu}\label{eq:TrGamma5-1}\\
    &&\Tr(\gamma_5\gamma^\mu\gamma^\nu\gamma^\rho\gamma^\sigma)=&2\imath(\delta^{\mu\nu}\epsilon^{\rho\sigma}-\delta^{\mu\rho}\epsilon^{\nu\sigma}+\delta^{\mu\sigma}\epsilon^{\nu\rho}\nonumber\\
    &&&+\delta^{\nu\rho}\epsilon^{\mu\sigma}-\delta^{\nu\sigma}\epsilon^{\mu\rho}+\delta^{\rho\sigma}\epsilon^{\mu\nu})\label{eq:TrGamma5-2}
  \end{align}
\end{subequations}
where $\bbbone$ is the identity in the Clifford algebra while the trace of an odd number of $\gamma$'s vanishes. In \eqref{eq:GammaFormulae}, the Levi-Civita symbol $\epsilon_{\mu\nu}$ satisfies $\epsilon_{01}=+1$. We are now in position to evaluate the one-loop contributions to the two- and four-point functions.\par
\section{Calculation of the Correlation Functions}
\subsection{The Two-point Function}
\begin{figure}
  \centering
  \includegraphics[scale=1]{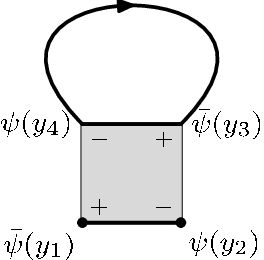}
  \caption[A Tadpole]{\footnotesize{A typical tadpole obtained from the contraction between Fermion fields occurring within one 
  operator 
  ${\cal{J}}$. The other tadpole of the same type is obtained through the substitution $y_1\leftrightarrow y_3$, 
  $y_2\leftrightarrow y_4$. Both diagrams give rise to equal contributions to the amputated two-point Green function.}}
  \label{fig:tadpole}
  \end{figure}

\begin{figure}
\centering
\includegraphics[scale=1]{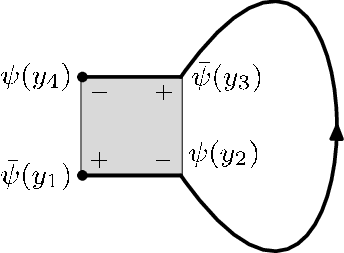}
  \caption[An Other Tadpole]{\footnotesize{A tadpole obtained from the contraction between Fermion fields taking place between 
  two operators 
  ${\cal{J}}$, the other one being obtained through $y_1\leftrightarrow y_3$, $y_2\leftrightarrow y_4$.}}
  \label{fig:tadpole2}
\end{figure}

Within the orientable non-commutative Gross Neveu model, the two-point Function receives contributions at the one-loop level from tadpole diagrams. There are obviously two types of tadpole diagrams depending whether the contraction between $\psi$ and ${\bar{\psi}}$ is performed within one among the two operators ${\cal{J}}$ defined in \eqref{eq:courants} that forms the whole vertex or the contraction takes place between the two operators ${\cal{J}}$. Typical tadpoles are depicted on the figures \ref{fig:tadpole} and \ref{fig:tadpole2}. 
The diagram on fig.~\ref{fig:tadpole} corresponds to a tadpole of the first type mentionned above. The corresponding amputated amplitude is given by
\begin{align}
  \cA_{1}=&-\sum_{A=1}^3\frac{g_A}{4\pi^2\theta^2}\int {\cal{D}}(y)\,
  {\bar{\psi}}(y_1)\Gamma^A\psi(y_2)\Tr(C(y_4,y_3)\Gamma^A)\label{eq:ATadpole1}
\end{align}
where the trace runs over spinor and color indices. It can be easily realized that the contribution from the other tadpole diagram, obtained through the substitution $y_1\leftrightarrow y_3$, $y_2\leftrightarrow y_4$ is equal to the one given in \eqref{eq:ATadpole1} which reflects the invariance of the phase factor in the vertex kernel \eqref{eq:InteractionKernel} under cyclic permutation. Therefore, both diagrams are taken into account simply by multiplying  the RHS of \eqref{eq:ATadpole1} by $2$.

In the same way, the figure~\ref{fig:tadpole2}
represents a typical tadpole of the second type mentioned above. The corresponding amplitude can be written as
\begin{align}
  \cA_{2}=\sum_{A=1}^3\frac{g_A}{4\pi^2\theta^2}\int{\cal{D}}(y)\,{\bar{\psi}}(y_1)\Gamma^AC(y_2,y_3)\Gamma^A\psi(y_4)\label{eq:ATadpole2}
\end{align}
while the amplitude stemming from the other tadpole diagram obtained from $y_1\leftrightarrow y_3$, $y_2\leftrightarrow y_4$ is equal to the RHS of \eqref{eq:ATadpole2} so that again taking into account both diagrams amounts to multiply the RHS of \eqref{eq:ATadpole2} by a factor $2$.

By further making use of \eqref{eq:Propa}, \eqref{eq:interaction} and the useful identity
\begin{align}
  \int\frac{d^2x}{(2\pi)^2}\,e^{\imath\alpha x\wedge z}=&\frac{\theta^2}{4\alpha^2}\delta(z),\label{eq:DeltaIdentity}
\end{align}
it can be easily seen that two among the four integrals over the space variables $y_i$'s can be explicitely performed thanks to delta functions so that \eqref{eq:ATadpole1} and \eqref{eq:ATadpole2} can be reexpressed as
\begin{align}
  {\cal{A}}_1=&-\sum_{A=1}^3\frac{g_A}{4(1+\Omega)^2}\int d^2y_1d^2y_2\,\delta(y_1-y_2)\int_0^\infty dt\,{\bar{\psi}}(y_1)\Gamma^A\psi(y_2)\Tr(\Cb(t;y_{1}-y_{2})\Gamma^A),\\
  {\cal{A}}_{2}=&\sum_{A=1}^3\frac{g_A}{4(1-\Omega)^2}\int d^2y_1d^2y_4\,\delta(y_1-y_4)\int_0^\infty dt\,{\bar{\psi}}(y_1)\Gamma^A\Cb(t;y_{1}-y_{4})\Gamma^A\psi(y_4),
  \intertext{which, upon integrating over the remaining delta functions, reduce respectively to}
  {\cal{A}}_1=&-\sum_{A=1}^3\frac{g_A}{4(1+\Omega)^2}\int d^2y dt\,{\bar{\psi}}(y)\Gamma^A\psi(y)\Tr(\Cb(t;0)\Gamma^A),\label{eq:ATad1Int}\\
  {\cal{A}}_{2}=&\sum_{A=1}^3\frac{g_A}{4(1-\Omega)^2}\int d^2y dt\,{\bar{\psi}}(y)\Gamma^A\Cb(t;0)\Gamma^A\psi(y).\label{eq:ATad2Int}
  \intertext{Relation \eqref{eq:ATad1Int} combined with \eqref{eq:PropaX} yields}
  {\cal{A}}_1=&-\sum_{A=1}^3\frac{m\Omega g_A}{4\pi\theta(1+\Omega)^2}\int d^2y dt\,{\bar{\psi}}(y)\Gamma^A\psi(y)\,e^{-tm^2}\big(\coth(2\Ot t)\Tr(\Gamma^A)-\Tr(\gamma_5\Gamma^A)\big)\label{eq:ATad1int2}
\end{align}
Then, by further inspecting the remaining integrals over the Schwinger parameter $t$, it is easy to see that the second term in \eqref{eq:ATad1int2} gives rise to a finite contribution and can therefore be ignored in the present analysis while the first term is UV logarithmically divergent as it can be realized from
\begin{align}
    I_1\defi\int_0^\infty dt\,e^{-m^2t}\coth(2\Ot t)=&\lim_{\epsilon\to 0}(-\frac{\theta}{4\Omega}\log\epsilon)+\dots 
\end{align}
where the ellipses denote finite contributions. Finally, since $\Tr(\Gamma^A)$ is non vanishing only when $\Gamma^A=\bbbone$, \eqref{eq:ATad1int2} yields
\begin{align}
  \cA_{1}=&-\frac{m\Omega g_1}{2\pi\theta(1+\Omega)^2} I_1\int d^2y\,{\bar{\psi}}(y)\psi(y)+\dots\label{eq:ATad1fin}
\end{align}
where again the ellipses denote finite contributions. In a similar way, we find that the logarithmically diverging part of \eqref{eq:ATad2Int} can be written as
\begin{align}
    \cA_{2}=&\frac{m\Omega}{4\pi\theta(1-\Omega)^2} I_1(g_1-2g_2+g_3)\int d^2y\,{\bar{\psi}}(y)\psi(y)+\dots\label{eq:ATad2fin}
\end{align}
Notice that the finite parts of \eqref{eq:ATad1fin} and \eqref{eq:ATad2fin} both involve a term equal to\\
$\delta\mu_i\int dy\,{\bar{\psi}}(y)\gamma_5\psi(y)$, $i=1,2$ with $\delta\mu_1=\frac{\Omega g_3}{2\pi m\theta(1+\Omega)^2}$ and $\delta\mu_2=-\frac{\Omega(g_1+2g_2+g_3)}{4\pi m\theta(1-\Omega)^2}$.

\subsection{The Four-point Function}
\begin{figure}[!hbt]
  \centering
  \includegraphics[scale=1]{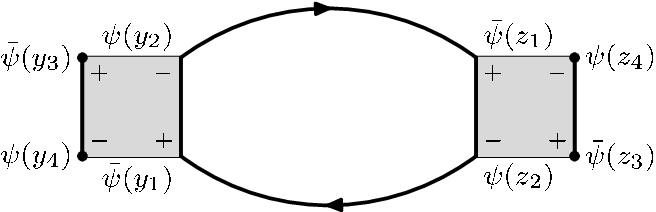}
  \caption[A Bubble]{\footnotesize{A UV logarithmically divergent planar diagram contributing to the $\beta$-functions, related to 
  the amplitude 
  \eqref{eq:ABubb1}.}}
  \label{fig:bubble1}
\end{figure}

\begin{figure}[!hbt]
  \centering
  \includegraphics[scale=1]{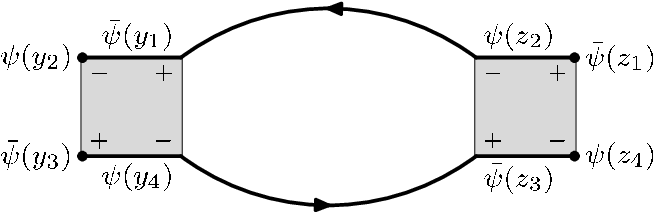}
  \caption[A Second Bubble]{\footnotesize{The other one-loop planar diagram contributing to the $\beta$-functions, related to the 
  amplitude 
  \eqref{eq:ABubb2}.}}
  \label{fig:bubble2}
\end{figure}

\begin{figure}[!hbt]
  \centering
  \includegraphics[scale=1]{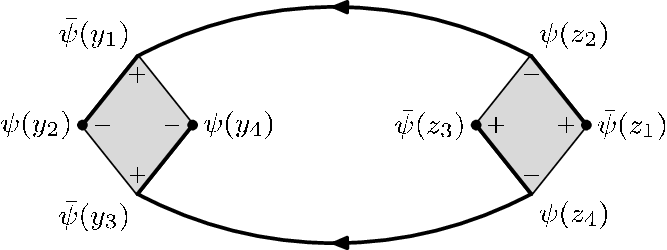}
  \caption[A Broken Bubble]{\footnotesize{The broken-face diagram corresponding to the amputated amplitude \eqref{glop}.}}
  \label{fig:BrokenBubble}
\end{figure}

\begin{figure}[!hbt]
\centering
\includegraphics[scale=1]{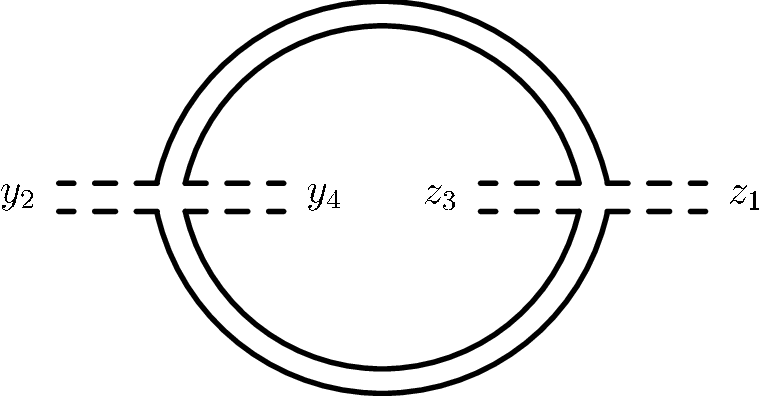}
\caption[A ribbon diagram]{\footnotesize{Ribbon diagram corresponding to the amputated amplitude \eqref{glop}, which is equivalent to the broken-face diagram depicted on the figure \ref{fig:BrokenBubble}}}
\label{fig:ribbondiag}
\end{figure}

The whole set of diagrams contributing to the $4$-point Function can be conveniently determined by finding all the possible ways to draw two contractions among the spinor fields involved in the correlation function 
\begin{align}
  \sum_{A,B}\frac{g_Ag_B}{16\pi^4\theta^4}\int{\cal{D}}(y){\cal{D}}(z)\,\langle 
  0|{\bar{\psi}}(y_1)\Gamma^A\psi(y_2){\bar{\psi}}(y_3)\Gamma^A\psi(y_4){\bar{\psi}}(z_1)\Gamma^B\psi(z_2){\bar{\psi
  }}(z_3)\Gamma^B\psi(z_4)|0\rangle\label{eq:4pt1}
\end{align}
while forbidding the graphs involving vacuum-vacuum subdiagrams. By fixing the first contraction to occur between ${\bar{\psi}}(y_1)$ and $\psi(z_2)$, one easily finds that the remaining contraction can be built from $5$ different ways in \eqref{eq:4pt1} generating $5$ different diagrams. These diagrams can be classified into two different types \cite{GrWu03-1} among which only the {\it{planar regular}} diagrams \cite{SIMPLIF}, plagued with UV logarithmic divergences, are relevant for the computation of the $\beta$-functions \cite{RenNCGN05}. These diagrams are depicted on the figures \ref{fig:bubble1} and \ref{fig:bubble2}. At this level, one comment is in order. Recall that the power-counting of a non-commutative field theory depends on the topology of its Feynman diagrams. These ones may be equivalently represented by ribbon diagrams. For example, the graph of the figure \ref{fig:BrokenBubble} corresponds also to the figure \ref{fig:ribbondiag}. From the ribbon representation, one can easily compute the genus $g$ (through the Euler characteristic) and the number of broken faces $B$ (defined as the number of faces to which external legs belong). For the non-commutative $\Phi^{4}$ theory \cite{GrWu03-1}, the superficial degree of convergence is
\begin{align}
\omega =& N-4+8g+4(B-1) \label{eq:counting}
\end{align}
with $N$ the number of external legs. From \eqref{eq:counting}, one infers that the only divergent diagrams have $g=0$ and $B=1$ (the {\it{planar regular}} ones) which are therefore the diagrams relevant for the calculation of the $\beta$ functions. In the Gross-Neveu case \cite{RenNCGN05}, the power-counting is slightly more complicated but the same conclusion holds.\par
The amputated amplitudes corresponding to the figures \ref{fig:bubble1} and \ref{fig:bubble2} are given respectively by
\begin{align}
  {\cal{B}}_{1}=&-\sum_{A,B}\frac{g_Ag_B}{16\pi^4\theta^4} \int{\cal{D}}(y){\cal{D}}(z)\,\Tr\big(C(z_2,y_1)\Gamma^AC(y_2,z_1)\Gamma^B\big){\bar{\psi}}(y_3)\Gamma^A\psi(y_4){\bar{\psi}}(z_3)\Gamma^B\psi(z_4),\label{eq:ABubb1}\\
  {\cal{B}}_{2}=&\sum_{A,B}\frac{g_Ag_B}{16\pi^4\theta^4}\int{\cal{D}}(y){\cal{D}}(z)\,
  {\bar{\psi}}(y_3)\Gamma^AC(y_4,z_3)\Gamma^B\psi(z_4){\bar{\psi}}(z_1)\Gamma^BC(z_2,y_1)\Gamma^A\psi(y_2).\label{eq:ABubb2}
\end{align}

The other type of one-loop diagrams, namely the broken-face diagrams, are UV finite \cite{RenNCGN05,GrWu03-1} and can therefore be ignored in the present analysis. For instance, one obtains the broken-face diagram depicted on the figure~\ref{fig:BrokenBubble}. Its corresponding amputated amplitude is given by
\begin{align}
  {\cal{B}}_{BF}=&\sum_{A,B}\frac{g_Ag_B}{16\pi^4\theta^4}\int{\cal{D}}(y){\cal{D}}(z)\,{\bar{\psi}}(z_1)
  \Gamma^BC(z_2,y_1)\Gamma^A\psi(y_2){\bar{\psi}}(z_3)\Gamma^BC(z_4,y_3)\Gamma^A\psi(y_4) \label{glop}
\end{align}
and can be explicitely verified to be UV finite.

We now extract the potentially diverging part of \eqref{eq:ABubb1} relevant for the computation of the $\beta$-functions. First, integrating over $y_2$ and $z_2$ and making use of \eqref{eq:Cbar} and the relation \eqref{eq:DeltaIdentity}, we find that \eqref{eq:ABubb1} can be cast into the form
\begin{subequations}
  \begin{align}
    {\cal{B}}_{1}=&\int d^2x_1d^2x_2d^2x_3d^2x_4\,{\cal{F}}_1(x_1,x_2;x_3,x_4)
    \intertext{with}
    {\cal{F}}_{1}=&-\sum_{A,B}\frac{g_Ag_B}{16\pi^2\theta^2(1+\Omega)^2}\delta(Y_1+Z_1)e^{-\imath(x_1\wedge x_2+x_3\wedge x_4)}{\bar{\psi}}(x_1)\Gamma^A\psi(x_2){\bar{\psi}}(x_3)\Gamma^B\psi(x_4)\nonumber\\
    &\times\int_0^\infty dt_1dt_2\int d^2u\,\Tr\big(\Cb(t_1;u+Z_1)\Gamma^A\Cb(t_2;Y_1-u)\Gamma^B\big)e^{\frac{\imath}{2}(1-\Omega)u\wed(Y_{1}-Z_{1})}\label{eq:ABubb1int1}
  \end{align}
\end{subequations}
where we have defined $Y_1=x_1-x_2$ and $\ Z_1=x_3-x_4$. Then, by further making use of the properties for the traces of products of $\gamma$ matrices given in \eqref{eq:GammaFormulae} combined with \eqref{eq:PropaX} and \eqref{eq:ExpGamma5}, we find that \eqref{eq:ABubb1int1} can be rewritten as
\begin{align}
    {\cal{F}}_{1}=&-\sum_{A,B}\frac{g_Ag_B\Omega^2\Ot^2}{16\pi^4\theta^4(1+\Omega)^2}\delta(Y_1+Z_1)e^{-\imath(x_1\wedge x_2+x_3\wedge x_4)}{\bar{\psi}}(x_1)\Gamma^A\psi(x_2){\bar{\psi}}(x_3)\Gamma^B\psi(x_4)\label{eq:ABubb1int2}\\
    &\times \int dt_1dt_2d^2X\,e^{-m^2(t_1+t_2)}(\kc_{1}\kc_{2})^{2}e^{-\frac{\Ot}{2}(\kc_{1}+\kc_{2})X^2-\frac{\imath}{2}(1-\Omega)X\wed(Y_{1}-Z_{1})}\Tr({\slashed{X}}\Gamma^A{\slashed{X}}\Gamma^B)+\dots\nonumber
\end{align}
in which $\kc_i=\coth(2\Ot t_i),\, i=1,2$ and the ellipses denote finite contributions. Finally, from the explicit computation of the trace in \eqref{eq:ABubb1int2}, one easily infers that the only non vanishing contributions are those where $\Gamma_A$ and $\Gamma_B$ are equal to each other, leading to ($X^2=X_\mu X^\mu$)
\begin{subequations}
  \begin{align}
    {\cal{B}}_{1}=&-\frac{N\Omega^2\Ot^2}{8\pi^2\theta^2(1+\Omega)^2}\int dx dX\,\cV(X)\Big(g_{1}^{2}X^{2}\psib\star\psi\star\psib\star\psi-g_{3}^{2}X^{2}\psib\star\gamma_{5}\psi\star\psib\star\gamma_{5}\psi\nonumber\\
    &\qquad-g_{2}^{2}(2X^{\mu}X_{\mu}-X^{2})\psib\star\gamma^{\mu}\psi\star\psib\star\gamma^{\mu}\psi \Big)(x)+\dots,\label{eq:ABubb1int3}\\
    \cV(X)=&\int_{0}^{\infty}dt_{1}dt_{2}\,e^{-m^2(t_1+t_2)}(\kc_{1}\kc_{2})^{2}e^{-\frac{\Ot}{2}(\kc_{1}+\kc_{2})X^2}.\label{eq:V(X)}
  \end{align}
\end{subequations}

A somehow similar analysis applied to \eqref{eq:ABubb2} permits one to extract the potentially diverging part relevant for the calculation of the $\beta$-functions. It takes the form
\begin{subequations}
  \begin{align}
    {\cal{B}}_{2}=&\int d^2x_1d^2x_2d^2x_3d^2x_4\,{\cal{F}}_2(x_1,x_2;x_3,x_4)
    \intertext{with}
    {\cal{F}}_2=&\sum_{A,B}\frac{g_Ag_B\Omega^2\Ot^2}{16\pi^4\theta^4(1-\Omega)^2}
    e^{-\imath(x_4\wedge x_1+x_2\wedge x_3)}\delta(Y_2+Z_2)\int d^2X\,\cV(X)\label{eq:ABubb2int1}\\
    &\times {\bar{\psi}}(x_1)\Gamma^A{\slashed{X}}\Gamma^B\psi(x_2)({\bar{\psi}}(x_3)\Gamma^B{\slashed{X}}\Gamma^A\psi(x_4)+\dots\nonumber
  \end{align}
\end{subequations}
in which $Y_2=x_4-x_1$, $Z_2=x_2-x_3$, $\cV(X)$ is given by \eqref{eq:V(X)} and the ellipses denote finite contributions. By repeated use of \eqref{eq:Gamma} and \eqref{eq:Gamma5}, the amplitude \eqref{eq:ABubb2int1} can be cast into the form
\begin{align}
  {\cal{B}}_2=&\frac{\Omega^2\Ot^2}{16\pi^2\theta^2(1-\Omega)^2}\int dx dX\,\cV(X)\label{eq:ABubb2int3}\\
  &\times\Big[(g_{1}^{2}+g_{3}^{2})(X^{\mu})^{2}\psib\star\gamma^{\mu}\psi\star\psib\star\gamma^{\mu}\psi+2g_{2}^{2}(X^{\mu})^{2}\psib\star\gamma^{\nu}\psi\star\psib\star\gamma^{\nu}\psi\nonumber\\
  &+2(g_{1}g_{2}+g_{2}g_{3})(X^{\mu})^{2}\big(\psib\star\psi\star\psib\star\psi+\psib\star\gamma_{5}\psi\star\psib\star\gamma_{5}\psi\big)\nonumber\\
  &+2g_{1}g_{3}(X^{\mu})^{2}\psib\star\gamma^{\mu+1}\psi\star\psib\star\gamma^{\mu+1}\psi\Big](x)+\dots\nonumber
\end{align}
where in the last line, $\mu\in\Z_{2}$.

It appears that ${\cal{B}}_1$ and ${\cal{B}}_2$ are plagued with UV logarithmic divergences as proven in \cite{RenNCGN05}. This can be easily verified by performing the integral over $X$ in \eqref{eq:ABubb1int3} and \eqref{eq:ABubb2int3} and then studying the behaviour of the resulting expressions when the Schwinger parameters become close to zero. Then using
\begin{align}
  \int_{0}^{\infty}dt_{1}dt_{2}\,e^{-(t_{1}+t_{2})m^{2}}\lbt\frac{\coth(\alpha t_{1})\coth(\alpha t_{2})}{\coth(\alpha t_{1})+\coth(\alpha t_{2})}\rbt^{2}=&\lim_{\epsilon\to 0^{+}}\frac{-\log\epsilon}{\alpha^{2}}+\cO(1),
\end{align}
the diverging parts of \eqref{eq:ABubb1int3} and \eqref{eq:ABubb2int3} are thus respectively given by
\begin{align}
  {\cal{B}}_1^{\text{div}}=&\lim_{\epsilon\to 0}-\frac{N\log\epsilon}{32\pi(1+\Omega)^2}\int dx \big(g_{1}^{2}\psib\star\psi\star\psib\star\psi-g_{3}^{2}\psib\star\gamma_{5}\psi\star\psib\star\gamma_{5}\psi\big)(x)\label{eq:ABubb1Div}\\
  {\cal{B}}_2^{\text{div}}=&\lim_{\epsilon\to 0}-\frac{\log\epsilon}{64\pi(1-\Omega)^2}\int dx \big(\lsb(g_{1}+g_{3})^{2}+4g_{2}^{2}\rsb\psib\star\gamma^{\mu}\psi\star\psib\star\gamma^{\mu}\psi\label{eq:ABubb2Div}\\
  &+4(g_{1}g_{2}+g_{2}g_{3})\big(\psib\star\psi\star\psib\star\psi+\psib\star\gamma_{5}\psi\star\psib\star\gamma_{5}\psi\big)\big)(x).\nonumber
\end{align}
\section{The {$\beta$} Functions}
\label{sec:beta-functions}

After having obtained the divergent parts of the relevant graphs at one loop, we are now in position to write down the beta functions. First of all note that the two-point function only enter the beta functions through the wave-function renormalisation. At one-loop order, only the tadpoles \eqref{eq:ATad1fin} and \eqref{eq:ATad2fin} contribute to the two-point function. It turns out that their divergent parts are exactly local and then only renormalise the mass. Then the beta functions at one-loop order are only computed from the four-point graphs. Notice that, as a byproduct, $\Omega$ is not renormalised to the one-loop order.

The interaction part of the effective action is
\begin{align}
  \Gamma_{\text{eff}}^{\text{int}}=&\int\frac{g_{1}}{4}\psib\star\psi\star\psib\star\psi+\frac{g_{2}}{4}\psib\star\gamma^{\mu}\psi\star\psib\star\gamma^{\mu}\psi+\frac{g_{3}}{4}\psib\star\gamma_{5}\psi\star\psib\star\gamma_{5}\psi\nonumber\\
  &+\frac{4}{2!}\Big(-\frac{N\log\epsilon}{32\pi(1+\Omega)^2}\int g_{1}^{2}\psib\star\psi\star\psib\star\psi-g_{3}^{2}\psib\star\gamma_{5}\psi\star\psib\star\gamma_{5}\psi\nonumber\\
  &-\frac{\log\epsilon}{64\pi(1-\Omega)^2}\int\lsb(g_{1}+g_{3})^{2}+4g_{2}^{2}\rsb\psib\star\gamma^{\mu}\psi\star\psib\star\gamma^{\mu}\psi\nonumber\\
  &+4(g_{1}g_{2}+g_{2}g_{3})\big(\psib\star\psi\star\psib\star\psi+\psib\star\gamma_{5}\psi\star\psib\star\gamma_{5}\psi\big)\Big).\label{eq:EffAction}
\end{align}
On the second line of \eqref{eq:EffAction}, the factor $1/2!$ comes from the expansion of the exponential of $S_{\text{int}}$ and the factor $4$ is just the number of Wick contractions leading to the considered graphs. By definition,
\begin{subequations}
  \begin{align}
    \beta_{i}=&\frac{d}{d(-\log\epsilon)}g_{i}(\{g_{jR}\})
    \intertext{where $g_{jR}$ stands for the renormalized constants. This gives}
    \beta_{1}=&-\frac{g_{1R}^{2}}{4\pi(1+\Omega)^{2}}-\frac{1}{2\pi(1-\Omega)^{2}}(g_{1R}g_{2R}+g_{2R}g_{3R})\\
    \beta_{2}=&-\frac{g_{2R}^{2}}{2\pi(1-\Omega)^{2}}-\frac{1}{8\pi(1-\Omega)^{2}}(g_{1R}+g_{3R})^{2}\\
    \beta_{3}=&\frac{g_{3R}^{2}}{4\pi(1+\Omega)^{2}}-\frac{1}{2\pi(1-\Omega)^{2}}(g_{1R}g_{2R}+g_{2R}g_{3R}).
  \end{align}
\end{subequations}

In \cite{RenNCGN05}, it has been proven that the action \eqref{eq:action} is renormalisable to all orders. From our one-loop computation, it seems that the stable manifold of parameters may well be reduced. Indeed from the expressions for the beta functions, it is clear that at one-loop, if $g_{1}=g_{3}=0$ then the interaction $g_{2}\psib\star\gamma^{\mu}\psi\star\psib\star\gamma^{\mu}\psi$ is stable. An other possibility is $g_{2}=0=g_{1}+g_{3}$ which is also stable. These two points are also true in the commutative case \cite{GROSSNEVEU}.

The table \ref{tab:CvsNC} summaries similarities and differences between the commutative and non-commutative Gross-Neveu and Thirring models. Remind that the commutative Gross-Neveu and Thirring models are given respectively by the following Lagrangians:
\begin{subequations}
  \begin{align}
  L_{\text{GN}}=&\int \psib(\ps+m)\psi-\lambda\psib\psi\psib\psi\\
  L_{\text{Th}}=&\int \psib(\ps+m)\psi-\lambda\psib\gamma^{\mu}\psi\psib\gamma_{\mu}\psi
\end{align}
\end{subequations}
\begin{table}
  \centering
  \begin{equation*}
        \begin{array}{|c|cc|cc|}
        \hline
        &\multicolumn{2}{|c|}{\rule[-3pt]{0pt}{15pt}\text{Commutative}}&\multicolumn{2}{|c|}{\text{Non-commutative}}\\
        \hline
        &\rule[-3pt]{0pt}{15pt}\text{Gross-Neveu}&\text{Thirring}&\text{Gross-Neveu}&\text{Thirring}\\
        \hline
        \raisebox{-5pt}{\includegraphics{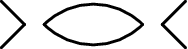}}&\rule[-3pt]{0pt}{18pt}N(\psib\psi)^{2}&0&N(\psib\psi)^{2}&0\\
        \raisebox{-8pt}{\includegraphics{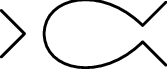}}&\rule[-3pt]{0pt}{18pt}-(\psib\psi)^{2}&0&\text{finite}&\text{finite}\\
        \raisebox{-8pt}{\includegraphics{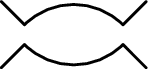}}&\rule[-3pt]{0pt}{18pt}0&0&(\psib\gamma^{\mu}\psi)^{2}&(\psib\gamma^{\mu}\psi)^{2}\\
        \hline
        &\text{stable}&\text{stable}&\text{unstable}&\text{stable}\\
        &\multicolumn{2}{|c|}{\text{asympt. free}}&&\text{asympt. free}\\
        &\multicolumn{2}{|c|}{\beta=0\text{ at }N=1}&&\beta>0\\
        \hline
      \end{array}
  \end{equation*}
  \caption{\footnotesize{Comparition between the Gross Neveu and Thirring models and their non-commutative counterparts.}}
  \label{tab:CvsNC}
\end{table}
\begin{figure}[!htb]
  \centering
  \includegraphics[scale=.45]{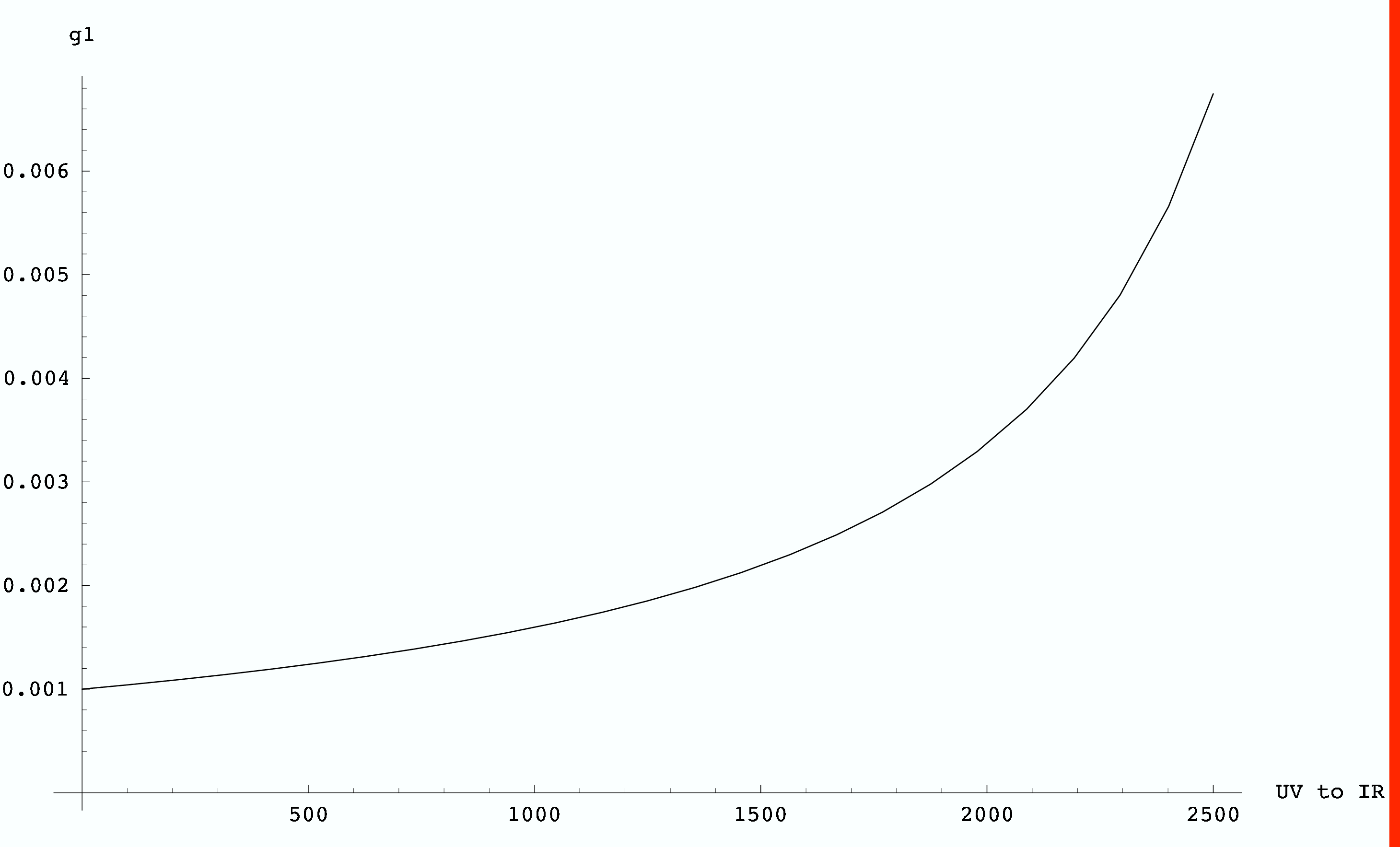}
  \caption{\footnotesize{UV Asymptotic Freedom for the model \eqref{eq:action} when $g_1=g_2=g_3$. The vertical (resp. 
  horizontal) axis represents $g_1$ (resp. $-\log\epsilon$). Notice that the UV region corresponds to the leftmost values. The units 
  are arbitrary.}}
 \label{fig:AF}
\end{figure}
In table \ref{tab:CvsNC}, the trajectories in the graphs correspond to contractions of spinor and color indices. For example the first and third graphs, in the non-commutative case, correspond respectively to figures \ref{fig:bubble1} and \ref{fig:bubble2}. On $\R^{2}$ both the Gross-Neveu and the Thirring models are stable (which means that no new interaction vertex is created by radiative corrections). On $\R^{2}_{\theta}$ the Gross-Neveu is not stable anymore. The third graph generates indeed a vertex of the form $\psib\star\gamma^{\mu}\psi\star\psib\star\gamma^{\mu}\psi$. This was not the case on $\R^{2}$ thanks to a compensation between two graphs. On $\R^{2}_{\theta}$ one of them is now finite (it has two broken faces). Note that the same phenomenon occurs for the second graph which becomes finite on $\R^{2}_{\theta}$.
A well-known fact about the commutative Gross-Neveu and Thirring model is their asymptotic freedom. This is also a feature of the non-commutative model \eqref{eq:action}. For $g_{1}=g_{2}=g_{3}$ (the bare values) or for $g_{1}=g_{3}=0$ or for $g_{2}=0=g_{1}+g_{3}$, we find that the model is asymptotically free in the UV region. A representative example is shown on the figure \ref{fig:AF} when $g_1=g_2=g_3$.

In \cite{GrWu04-2}, the one-loop beta function for the non-commutative $\Phi^{4}$ model has been computed. It was shown that for $\Omega<1$, the flow is bounded contrary to the commutative case where the theory is asymptotically free in the IR. Note that this is not the case for the non-commutative Gross-Neveu model. Moreover, at $\Omega=1$, the beta function of $\Phi^{4}$ vanishes at any order (asymptotically  in the UV region) \cite{DISSERT}. In the present case, the limit $\Omega\to 1^{-}$ is singular as it can be seen from equation e.g. \eqref{eq:ABubb2Div}. This singularity hides the actual behaviour of the $\beta$-functions at $\Omega=1$ so that further investigations are needed to determine whether or not vanishing $\beta$ can also be observed whenever $\Omega=1$, as it is the case for the $\Phi^{4}$ model. Finally recall that for $N=1$, the beta function of the equivalent Thirring and Gross-Neveu models on $\R^{2}$ vanishes. This feature is also lost on $\R^{2}_{\theta}$.
\vskip 1 true cm
\paragraph{Acknowledgement}Discussions with J. Magnen, V. Rivasseau and R. Wulkenhaar at various stages of this work are gratefully acknowledged.

\end{document}